\documentclass[%
twocolumn,
prd,
nofootinbib,
superscriptaddress
]{revtex4}
\usepackage{amsmath,amssymb,bm,float,mathrsfs,tabularx}
\usepackage[dvipdfmx]{graphicx}
\usepackage{dcolumn}
\usepackage{color}
\usepackage{hyperref}
\usepackage{comment}

  \newcommand{\beq}{\begin{equation}}
  \newcommand{\eeq}{\end{equation}}
  \newcommand{\al}[1]{\begin{align} #1 \end{align}}
  \newcommand{\bi}{\begin{itemize}}
  \newcommand{\ei}{\end{itemize}}
  \def\dd{\mathrm{d}}

  \def\pd{\partial}
  \newcommand{\ave}[1]{\left\langle #1 \right\rangle}

\begin{document}

\title{Multitracer technique for galaxy bispectrum \\
- An application to constraints on non-local primordial non-Gaussianities -}


\author{Daisuke Yamauchi}
\email[Email: ]{yamauchi"at"jindai.jp}
\affiliation{
Faculty of Engineering, Kanagawa University, Kanagawa, 221-8686, Japan
}
\author{Shuichiro Yokoyama}
\affiliation{
Department of Physics, Rikkyo University, Tokyo 171-8501, Japan
}

\author{Keitaro Takahashi}
\affiliation{
Faculty of Science, Kumamoto University, 2-39-1 Kurokami, Kumamoto 860-8555, Japan
}

\begin{abstract}
We explore the use of galaxy bispectra with multitracer technique as a possible probe of primordial non-Gaussianities.
We forecast future constraints on non-linearity parameters, $f_{\rm NL}^{\rm eq}$ and $f_{\rm NL}^{\rm orth}$, which respectively characterize the equilateral- and orthogonal-types primordial bispectra,
and show that the multitracer analysis would be effective with reducing the cosmic-variance noise
if the number density of galaxies is high enough.
We find that
the measurement of galaxy bispectrum by future galaxy surveys can reach the constraints 
on the non-local type primordial non-Gaussianities to the level severer than current one which has been obtained by cosmic microwave background observations.
\end{abstract}

\pacs{pacs}
\preprint{RUP-16-29}

\maketitle

\section{Introduction} 

Primordial non-Gaussianity (PNG), recently, has been expected to be one of the most informative fingerprints
of inflation and brings insights into the fundamental physics behind inflation.
Since different shapes of higher-order primordial spectra can be linked to different 
mechanism for generating non-Gaussian features of the primordial fluctuations, it would be interesting 
to constrain various types of PNG by precise cosmological measurements
such as cosmic microwave background (CMB) and large scale structure (LSS).

The amplitudes in the various shapes of the primordial bispectrum are basically 
characterized by three so-called non-linearity parameters
$f_{\rm NL}^{\rm local}$\,, $f_{\rm NL}^{\rm eq}$\,, and $f_{\rm NL}^{\rm orth}$,
which respectively correspond to the amplitudes of the {\it local}-~\cite{Komatsu:2001rj}, {\it equilateral}-~\cite{Creminelli:2005hu} , and {\it orthogonal}-type~\cite{Creminelli:2005hu}.
These are frequently considered as typical examples and are strongly motivated by inflationary models.
The current limits on these parameters have been obtained from the CMB temperature anisotropies and polarizations:
$f_{\rm NL}^{\rm local}=0.8\pm 5.0$\,, 
$f_{\rm NL}^{\rm eq}=-4\pm 43$\,, and
$f_{\rm NL}^{\rm orth}=-26\pm 21$ at $1\sigma$ statistical significance, respectively~\cite{Ade:2015ava,Ade:2013ydc}.
Although such as strict constraints on the non-linearity parameters have not been obtained from the observational data of LSS yet,
the spatial clustering behavior of the halos/galaxies on large scales is believed to be a powerful tool to probe the PNG.
One of the most distinctive effect of the PNG on the clustering of the galaxies is known as  {\it scale-dependent bias} (see e.g., \cite{Dalal:2007cu,Matarrese:2008nc}),
which is due to the non-linear coupling caused by PNG,
and it is expected to have a potential to reach $\sigma (f_{\rm NL}^{\rm local}) = {\mathcal O} (0.1 - 1)$ in future surveys.

It is, however, shown that the scale-dependent clustering property due to PNG
extracted from galaxy power spectrum is too weak to detect except for the local-type,
implying that the scale-dependent clustering is irrelevant for the non-local type PNG~\cite{Matsubara:2012nc}.
Then, the bispectrum of the biased objects such as halo/galaxy
has been considered as one of the useful observables to obtain constraints on the non-local type PNG.
Although the halo/galaxy bispectrum should be generated from the late-time non-linear gravitational evolution of the
density fluctuations, the contribution from the non-local type PNG would be dominant on larger scales
(see, e.g., \cite{Sefusatti:2007ih,Yokoyama:2013mta,Mizuno:2015qma,Hashimoto:2015tnv,Hashimoto:2016lmh}).
We then expect that the property of the scale-dependence of the galaxy bispectrum 
provides us the opportunity to probe not only the local but also non-local type PNG
with precise measurements of LSS in future.
However, the clustering analysis at large scales is limited due to cosmic-variance (CV), because of
the lack of enough independent measurements.
A possible way to reduce the CV noise is the use of multitracer technique~\cite{Seljak:2008xr,Hamaus:2011dq},
in which the availability of multiple tracers with different biases allows significant improvements
in the statistical errors.
The clustering analysis with multitracer technique has been previously studied only in the case of
the power spectrum for future galaxy surveys such as Euclid~\footnote{See http://www.euclid-ec.org} 
and SKA~\footnote{See http://www.skatelescope.org}, and 
the effect turns out to be indeed very effective and we can reach $\sigma (f_{\rm NL}^{\rm local})={\cal O}(1)$
\cite{Ferramacho:2014pua,Yamauchi:2014ioa,Yamauchi:2016ypt,Yamauchi:2015mja,Kitching:2015fra,Takahashi:2015zqa}
even when the horizon-scale effects are taken into account~\cite{Fonseca:2015laa,Alonso:2015sfa,Alonso:2015uua,Camera:2014bwa}.
However, the multitracer method for the higher moments has not been discussed in the literature.

In this paper, thus, we consider the galaxy bispectrum as the probe of the various types of PNG 
and derive the formulas for the monopole mode of the bispectrum 
from the redshift space distortion (RSD) generalized to multiple tracers to apply the multitracer method.
Based on the derived analysis tool, we then calculate the expected galaxy bispectrum and
the possibility to detect the PNG is discussed.


\section{Bispectrum with multiple tracers} 

The power spectrum, $P^{(g_1g_2)}(k)$, and bispectrum, $B^{(g_1g_2g_3)}(k_1,k_2,k_3)$, where we have used a index $g_i$ as a label of the tracer objects such as galaxies,
can be defined in terms of
the Fourier components of the number density field of the tracer objects, $\delta^{(g_i)}({\bm k})$, 
as 
\al{
	&\ave{\delta^{(g_1)}({\bm k})\delta^{(g_2)}({\bm k}')}
		=(2\pi )^3\delta^3_{\rm D}({\bm k}+{\bm k}')P^{(g_1g_2)}(k)
	\,,\\
	&\ave{\delta^{(g_1)}({\bm k}_1)\delta^{(g_2)}({\bm k}_2)\delta^{(g_3)}({\bm k}_3)}
	\notag\\
	&\quad\quad
		=(2\pi )^3\delta^3_{\rm D}({\bm k}_1+{\bm k}_2+{\bm k}_3)B^{(g_1g_2g_3)}(k_1,k_2,k_3)
	\,,
}
where $\delta^3_{\rm D}$ is a 3-dimensional Dirac's delta function.
Even if the initial condition for density fluctuations is assumed to be Gaussian,
the nonlinear gravitational evolution naturally induces the non-negligible non-Gaussianity.
In large-scale limit where the scale of interest is much larger than the typical scale
of the collapsed objects, the bispectrum of the tracer objects can be decomposed 
into several parts~\cite{Yokoyama:2013mta}.
Hereafter we focus on the dominant contributions on large scales
and let us consider the contributions from the gravitational evolution and
the primordial bispectrum, $B_{\rm grav}$ and $B_{\rm bis}$, which can be simply written as~\cite{Sefusatti:2007ih,Yokoyama:2013mta}
\al{
	&B_{\rm grav}^{(g_1g_2 g_3)}
		=\frac{1}{6}
			\biggl[
				b_1^{(g_1)}b_1^{(g_2)}\left( b_2^{(g_3)}+2b_1^{(g_3)}F_2({\bm k}_1,{\bm k}_2)\right)
	\notag\\
	&\quad\quad\quad
				+\left( g_i\,{\rm perm}\right)
		\biggr]P_{\rm L} (k_1)P_{\rm L} (k_2)
			+\left( k_i\,{\rm perm}\right)
	\,,\\
	&B_{\rm bis}^{(g_1g_2g_3)}
		=b_1^{(g_1)}b_1^{(g_2)}b_1^{(g_3)}B_{\rm L}(k_1,k_2,k_3)
	\,,
}
where 
$P_{\rm L}(k)$ and $B_{\rm L} (k_1, k_2, k_3)$ are, respectively, a power spectrum and a bispectrum for the linear density field, $\delta_{\rm L}$.
The linear density field can be related to  the primordial curvature perturbations $\Phi$ through
the Poisson equation as $\delta_{\rm L}({\bm k},z)={\cal M}(k;z)\Phi ({\bm k})$ with
${\cal M}(k;z)=2D_+(z)k^2T(k)/3H_0^2\Omega_{{\rm m},0}$\,, 
where $D_+(z)\,,T(k)$ represent the linear growth rate, matter transfer function normalized to unity at large scale~\cite{Eisenstein:1997ik},
respectively. Based on this expression, 
we can rewrite $P_{\rm L}$ and $B_{\rm L}$ as $P_{\rm L}(k)={\cal M}^2(k)P_\Phi (k)$ and 
$B_{\rm L}(k_1,k_2,k_3)={\cal M}(k_1){\cal M}(k_2){\cal M}(k_3)B_\Phi (k_1,k_2,k_3)$.
Furthermore, $F_2$ corresponds to the second-order kernel of standard perturbation theory and
$b_1^{(g_i)}$ and $b_2^{(g_i)}$ denote the linear and nonlinear bias parameters for the $g_i$-th tracer object,
respectively.
We note that when deriving the bispectrum shown above we have considered the perturbative expansion
up to the tree-level order. Here we simply neglect the higher-order loop contributions because they are
expected to be not so significant at large scales~\cite{Yokoyama:2013mta}.


The observed power- and bi-spectra from redshift surveys are distorted by the radial motion of galaxies.
In order to consider the RSD, we assume that the higher-order contributions are neglected.
Then the leading-order expression for the galaxy power-/bi-spectrum 
with the redshift space distortion is given by~\cite{Scoccimarro:1999ed}
\al{
	&P_{s}^{(g_1g_2)}
		=Z_1^{(g_1)}({\bm k})Z_1^{(g_2)}({\bm k})P_{\rm L}(k)
	\,,\\
	&B_{{\rm grav},s}^{(g_1g_2 g_3)}
		=\frac{1}{6}
			\biggl[
				2Z_1^{(g_1)}({\bm k}_1)Z_1^{(g_2)}({\bm k}_2)Z_2^{(g_3)}({\bm k}_1,{\bm k}_2)
	\notag\\
	&\quad\quad\quad
				+\left( g_i\,{\rm perm}\right)
		\biggr]P_{\rm L} (k_1)P_{\rm L} (k_2)
			+\left( k_i\,{\rm perm}\right)
	\,,\\
	&B_{{\rm bis},s}^{(g_1g_2g_3)}
		=Z_1^{(g_1)}({\bm k}_1)Z_1^{(g_2)}({\bm k}_2)Z_1^{(g_3)}({\bm k}_3)
		B_{\rm L} (k_1,k_2,k_3)
	\,,
}
where the linear- and second- perturbation theory kernels $Z_n^{(g)}$ are defined as
\al{
	&Z_1^{(g)}({\bm k})
		=b_1^{(g)}+f\mu^2
	\,,\\
	&Z_2^{(g)}({\bm k}_1,{\bm k}_2)
		=\frac{1}{2}b_2^{(g)}+b_1^{(g)}F_2({\bm k}_1,{\bm k}_2)
	\,.
}
with $\mu$ being the cosine of angle to the line of sight.
We have dropped the contributions from the second-order velocity kernel and velocity dispersion.
Since the distorted spectra are rather complicated, we instead deal only with a spherically averaged power
and bispectrum in subsequent analysis, following Refs.~\cite{Sefusatti:2007ih,Scoccimarro:1999ed,Sefusatti:2006pa}.
To do so, we first derive the monopole contributions from RSD generalized to multiple tracers.
By averaging over the angle between ${\bm k}_1$ and ${\bm k}_2$ and dropping the angle-dependent term, 
we find the averaged power- and bi-spectra in redshift space are respectively given as
\al{
	&P_{0}^{(g_1g_2)}=a^{(g_1g_2)}_{0,{\rm pow}}\,P^{(g_1g_2)}
	\,,\\
	&B_{{\rm grav},0}^{(g_1g_2g_3)}
		=\frac{1}{6}
			\biggl[
				a_{0,{\rm bis}}^{(g_1g_2)}b_1^{(g_1)}b_1^{(g_2)}\left( b_2^{(g_3)}+2b_1^{(g_3)}F_2({\bm k}_1,{\bm k}_2)\right)
	\notag\\
	&\quad\quad\quad
				+\left( g_i\,{\rm perm}\right)
		\biggr]P_{\rm L} (k_1)P_{\rm L} (k_2)
			+\left( k_i\,{\rm perm}\right)
	\,,\\
	&B_{{\rm bis},0}^{(g_1g_2g_3)}
		=\frac{1}{6}\left( a_{0,{\rm bis}}^{(g_1g_2)}+(g_i\,{\rm perm})\right) B^{(g_1g_2g_3)}
	\,,	
}
where the monopole redshift-space correction are
\al{
	&a^{(g_1g_2 )}_{0,{\rm pow}}
		=1+\frac{1}{3}\left(\beta^{(g_1)}+\beta^{(g_2)}\right)+\frac{1}{5}\beta^{(g_1)}\beta^{(g_2)}
	\,,\\
	&
		a_{0,{\rm bis}}^{(g_1g_2)}=\left( 1+\frac{1}{3}\beta^{(g_1)}\right)\left( 1+\frac{1}{3}\beta^{(g_2)}\right)
	\,,
}
with $\beta^{(g_i)}:=f/b_1^{(g_i)}$. The function $f$ is the linear growth rate, which is
defined by the logarithmic derivative of the linear density field with respect to 
the logarithmic of scale factor.
The above expressions are the multitracer generalization of the formula for the single tracer, given in Refs.~\cite{Sefusatti:2007ih,Scoccimarro:1999ed,Sefusatti:2006pa}.


The galaxy bias parameters can be calculated from the dark matter halo bias,
if we assume that galaxies are formed in dark matter halos.
To evaluate the galaxy biases, we shall use the halo bias parameters $b_\ell^{\rm h}(M,z)$ 
given in \cite{Scoccimarro:2000gm}, 
the Sheth-Tormen mass function $\dd n/\dd M$ \cite{Sheth:1999mn},
and the mean number of galaxies per halo of a given mass $M$, 
namely the halo occupation distributions $\ave{N}_M$~\cite{Tinker:2004gf}
with fitting parameters given in \cite{Conroy:2005aq}.
The explicit form of the linear and nonlinear halo bias parameters $b_1^{\rm h}$ and $b_2^{\rm h}$
are given by
\al{
	&b_1^{\rm h}(M,z)=1+\epsilon_1+E_1
	\,,\\
	&b_2^{\rm h}(M,z)=\frac{8}{21}\left(\epsilon_1+E_1\right)+\epsilon_2+E_2
	\,,
} 
where 
\al{
	&\epsilon_1=\frac{q\nu^2 -1}{\delta_{\rm c}}
	\,,\ \ \ 
	\epsilon_2=\frac{q\nu^2}{\delta_{\rm c}}\frac{q\nu^2-3}{\delta_{\rm c}}
	\,,\\
	&E_1=\frac{2p/\delta_{\rm c}}{1+(q\nu^2)^p}
	\,,\ \ \ 
	\frac{E_2}{E_1}=\frac{1+2p}{\delta_{\rm c}}+2\epsilon_1
	\,,
}
with $\nu=\delta_{\rm c}/\sigma (M,z)$\,, $\delta_{\rm c}=1.686$\,, $p=0.3$\,, and $q=0.707$\,.
Moreover, we split whole galaxy samples into some mass-divided subsamples
for each redshift bin to apply the multitracer technique.
The averaged number density of galaxies for the $g_i$-th mass bin, $M_{(g_i)}<M<M_{(g_{i+1})}$\,, is given by
\al{
	n_{(g_i)}=\int_{M_{\rm min}}\dd M\,\frac{\dd n}{\dd M}\,S_{(g_i)}\ave{N}_M\,,
}
where $S_{(g_i)}$ and $M_{\rm min}$ represent the selection function and 
the minimum mass above which we find a central galaxy in the halo, respectively.
In this paper we will simply take the top-hat form of the selection function as
$S_{(g_i)}=\Theta\left( M-M_{(g_i)}\right)\Theta\left( M_{(g_{i+1})}-M\right)$.
We find $M_{\rm min}$ from the total galaxy number density
\al{
	n_{\rm g}=\int_{M_{\rm min}}\dd M\,\frac{\dd n}{\dd M}\ave{N}_M
	\,,
}
for a given $n_{\rm g}$.
With these, we calculate the galaxy bias parameters for the $g_i$-th mass bin 
from the large scale expression
\al{
	b_\ell^{(g_i)}
		=\frac{1}{n_{(g_i)}}\int_{M_{\rm min}}\dd M\,\frac{\dd n}{\dd M}\,S_{(g_i)}\,b^{\rm h}_\ell\,\ave{N}_M~(\ell = 1,2)
	\,.
}


\section{Fisher analysis} 

Before going to the detail evaluation of the Fisher matrix, we shall discuss 
about the analytic understanding of the merit of the multitracer technique in the galaxy bispectrum.
Just for simplicity, 
first, we focus only on the galaxy power- and bi-spectra from a single mode with wavelength $k$.
Moreover, we drop the contributions of the galaxy bispectrum of general triangle configurations 
except for the equilateral one and the effect of RSD.
Then, the Fisher matrix for a given wavelength $k$ is defined as
\al{
\widetilde F_{\alpha\beta}(k)=\frac{\pd{\bm B}^{\rm eq}}{\pd\theta^\alpha}\cdot [\widetilde{\rm C}^{-1}({\bm B}^{\rm eq},{\bm B}^{\rm eq})]\cdot\frac{\pd{\bm B}^{\rm eq}}{\pd\theta^\beta}\,,
}
where ${\bm B}^{\rm eq}\equiv\{ B^I(k,k,k)\}$, $I$ runs over the combination of the mass bins,
and $\theta^\alpha$ are free parameters.
Since we are interested only in the asymptotic behavior of the Fisher matrix, 
the covariance matrix we consider here is assumed to be 
\al{
\widetilde{\rm C}_{IJ}=\frac{1}{6}(\widehat P^{(g_1g_1^\prime)}\widehat P^{(g_2g_2^\prime )}\widehat P^{(g_3g_3^\prime )}+({\rm perm}))\,,
}
with $\widehat P^{(g_ig_j^\prime )}\equiv P^{(g_ig_j^\prime)}+n_{(g_i)}^{-1}\delta_{g_ig_j^\prime}^{\rm K}$ 
denoting the galaxy power spectrum including the shot-noise contamination.
Here, $\delta^{\rm K}$ is a Kronecker's delta function.
Note that the equilateral limit of the galaxy bispectrum can be reduced to
the simple form:
\al{
	&B^{(g_1g_2g_3)}(k,k,k)
		=\left(
			b_1^{(g_1)}b_1^{(g_2)}\widetilde b_2^{(g_3)}
			+(g_i\ {\rm perm})
		\right) P_{\rm L}^2(k)
	\,,
}
where we have introduced the effective nonlinear bias parameter
including the contribution from PNG, which is defined as 
$\widetilde b_2^{(g)}:=(b_2^{(g)}+\frac{4}{7}b_1^{(g)})+b_1^{(g)}f_{\rm NL}/{\cal M}$\,.
We then perform the Fisher analysis to forecast the future constraint 
on the nonlinearity parameter $f_{\rm NL}$.
Since the explicit form of the constraint on $f_{\rm NL}$ is rather complicated, 
we will instead evaluate the error on the relative effective nonlinear bias parameter.
Let us assume that two tracers of underlying dark matter density field, that is, $g_i = 1, 2$.
In this case, $I$ runs over $4$ bins, namely $I=\{(111)\,,(112)\,,(122)\,,(222)\}$.
Introducing the relative linear and nonlinear bias parameters
$\alpha\equiv b_1^{(1)}/b_1^{(2)}$ and $\gamma\equiv\widetilde b_2^{(1)}/\widetilde b_2^{(2)}$, 
we can rewrite the galaxy bispectrum as
$B^{(222)}=:{\cal B}$\,,
$B^{(111)}=\alpha^2\gamma{\cal B}$\,,
$B^{(112)}=(\alpha^2+2\alpha\gamma ){\cal B}/3$\,,
and $B^{(122)}=(2\alpha +\gamma ){\cal B}/3$.
Furthermore, the covariance matrix elements are also written in terms
of the relative linear bias $\alpha$ and the stochasticity parameter $r$ as
 $\widehat P^{(22)}={\cal P}(1+X_2)$ \,,
$\widehat P^{(11)}={\cal P}(\alpha^2+X_1)$\,, and 
$\widehat P^{(12)}=r\alpha{\cal P}$ with $X_{g_i}=1/(n_{(g_i)}{\cal P})$.
Assuming little stochasticity ($r\to 1$) and computing the error on the parameter $\gamma$ 
in the small shot-noise limit ($X_{g_i}\to 0$),
we obtain the leading-order term of the unmarginalized $1\sigma$ error on the parameter $\gamma$ 
in the noise/power ratio for both tracers:
\al{
	\widetilde\sigma^2 (\gamma)
		\equiv\widetilde F_{\gamma\gamma}^{-1}
		\approx\frac{3{\cal P}^3}{{\cal B}^2}\Bigl\{ X_1+\alpha^2 X_2+\alpha^2\left( 1-r^2\right)\Bigr\}
	\,.
}
This implies that the error on the relative effective nonlinear bias $\gamma$ from a single mode can be much less
than unity if there is little stochasticity and the field is oversampled.
Therefore, it is expected that with the multitracer technique we could measure PNG
via the galaxy bispectrum without the CV noise, even if the type of PNG
is non-local such as equilateral and orthogonal.

\begin{figure}[t]
\includegraphics[width=65mm]{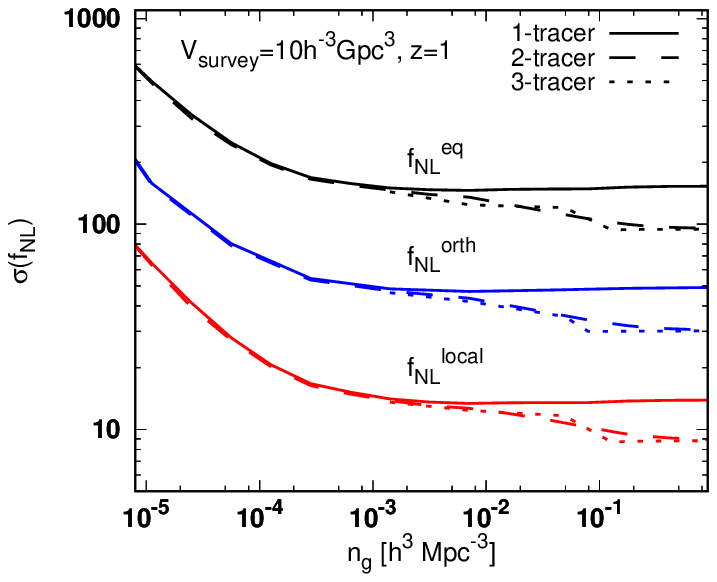}
\caption{
$1\sigma$ marginalized errors in 
as a function of comoving galaxy number density, 
$f_{\rm NL}^{\rm eq}$ (top), $f_{\rm NL}^{\rm orth}$ (middle) and $f_{\rm NL}^{\rm local}$ (bottom)
for survey with $V_{\rm survey}=10\,h^{-3}{\rm Gpc}^3$ and $z=1$.
}
\label{fig:plot_sigma_fNL_ng}
\end{figure} 

\begin{figure}[t]
\includegraphics[width=65mm]{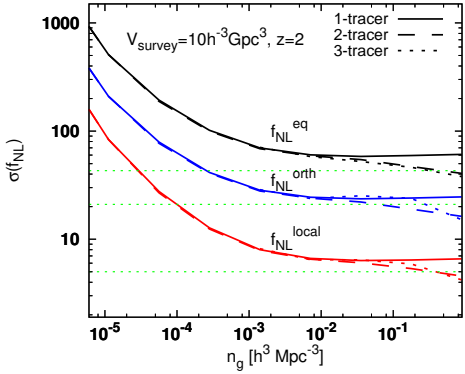}
\caption{
Same as Fig.~\ref{fig:plot_sigma_fNL_ng} but for $z=2$.
The green dotted lines represent the current limit from Planck~\cite{Ade:2015ava,Ade:2013ydc}.
}
\label{fig:plot_sigma_fNL_ng_z_2}
\end{figure} 

Let us numerically investigate the expected constraint on the non-linearity parameters
with the multitracer technique,
based on the full Fisher analysis.
In order to evaluate the expected future constraints, we calculate the Fisher matrix for the bispectrum,
which is obtained by summing over all possible triangular configurations.
The explicit expression is  given by \cite{Scoccimarro:2003wn}
\al{
	F_{\alpha\beta}
		=\sum_{k_1,k_2,k_3=k_{\rm min}}^{k_{\rm max}}\,
			\frac{\pd {\bm B}}{\pd\theta^\alpha}
			\cdot
			\bigl[
				{\rm C}^{-1}({\bm B},{\bm B})
			\bigr]
			\cdot
			\frac{\pd{\bm B}}{\pd\theta^\beta}
	\,,\label{eq:Fisher}
}
where ${\bm B}=\{ B_0^I({\bm k}_1,{\bm k}_2,{\bm k}_3)\}$, $I$ runs over the mass bins $(g_1g_2g_3)$,
and $\theta^\alpha$ are free parameters to be determined by observations.
The marginalized expected $1\sigma$ error on parameter $\theta^\alpha$ from the Fisher matrix \eqref{eq:Fisher}
 is estimated to be $\sigma (\theta^\alpha )=\sqrt{(F^{-1})_{\alpha\alpha}}$\,.
Assuming the Gaussian error covariance, we obtain the covariance matrix for multiple tracers as~\cite{Sefusatti:2007ih,Sefusatti:2006pa,Scoccimarro:2003wn}
\al{
	&{\rm C}_{IJ}
		=\frac{s_{\rm B}V_{\rm survey}}{36N_{\rm t}}
	\notag\\
	&\times
			\Bigl[
				\widehat P_0^{(g_1g_1^\prime)}(k_1)\widehat P_0^{(g_2g_2^\prime )}(k_2)\widehat P_0^{(g_3g_3^\prime )}(k_3)
		+({\rm perm})
			\Bigr]
	\,,
}
where $s_{\rm B}$ is the symmetric factor describing the number of a given bispectrum triangle 
($s_{\rm B}=6\,,2\,,$ and $1$ for equilateral, isosceles and general triangles, respectively)
and the quantity $N_{\rm t}=V_{\rm B}/k_{\rm F}^6$ denotes the total number of available triangles 
with $k_{\rm F}=2\pi /V_{\rm survey}^{1/3}$ and $V_B=8\pi^2 k_1k_2k_3(\Delta k)^3$ being the fundamental frequency
and the volume of the fundamental cell in Fourier space, respectively.
Here, $\widehat P_0^{(g_1g_2)}$ is the averaged redshift-space galaxy power spectrum including the shot-noise contamination given by
$\widehat P_0^{(g_ig_j^\prime )}(k)=P_0^{(g_ig_j^\prime )}(k)+n_{(g_i)}^{-1}\delta_{g_ig_j^\prime}^{\rm K}$\,.
In subsequent analysis we assume that both the frequency gap and the minimum wavelength 
coincide with the fundamental frequency, namely $k_{\rm F}=\Delta k=k_{\rm min}$.
Moreover, for the maximal wavelength we choose $k_{\rm max}=\pi /(2R_{\rm min})$ with $R_{\rm min}$ 
such that $\sigma (R_{\rm min},z)=0.5$ \cite{Sefusatti:2007ih}.
For instance, $k_{\rm max}=0.19\,[h\,{\rm Mpc}^{-1}]$ at $z=1$ and $k_{\rm max}=0.35\,[h\,{\rm Mpc}^{-1}]$ at $z=2$.
Throughout this paper, as our fiducial model we assume a $\Lambda$CDM cosmological model
with parameters: 
$\Omega_{{\rm m},0}=0.318$, $\Omega_{{\rm b},0}=0.0495$, $\Omega_{\Lambda ,0} =0.6817$,
$w=-1$, $h=0.67$, $n_{\rm s}=0.9619$, $k_0=0.05{\rm Mpc}^{-1}$, $\sigma_8=0.835$.

Let us consider two and three mass bins such that each mass bin has the same galaxy number density,
simply because the tightest constraint for $f_{\rm NL}$ is expected to be obtained
when the shot-noise contributions from all mass bins become comparable. 
In the case of the two (three) mass bins, we have the five (seven) parameters in the Fisher matrix analysis:
the four (six) bias parameters and nonlinearity parameter $f_{\rm NL}$.
The bias parameters are fully marginalized over when deriving constraint on $f_{\rm NL}$.
When forecasting each nonlinearity parameter, we neglect the other parameters.
The fiducial values of the nonlinearity parameters are set to zero and 
the fiducial values of the linear and nonlinear bias parameters are calculated for each redshift.

\begin{figure}[t]
\includegraphics[width=75mm]{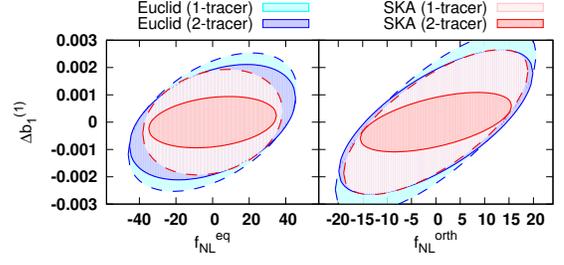}
\caption{
Forecast $1\sigma$ marginalized contours in $(f_{\rm NL}^{\rm eq},b_1^{(1)})$ (left),
$(f_{\rm NL}^{\rm orth},b_1^{(1)})$ (right) planes. 
}
\label{fig:fNL-b1}
\end{figure} 

To see the impact of the multitracer technique in the measurement of the galaxy bispectrum,
we show the expected marginalized $1\sigma$ statistical errors on the nonlinearity parameters 
in Figs.~\ref{fig:plot_sigma_fNL_ng} and \ref{fig:plot_sigma_fNL_ng_z_2}, marginalizing over the bias parameters.
The fiducial survey parameters are given by the survey volume $V_{\rm survey}=10\,h^3{\rm Mpc}^{-3}$ and 
the redshifts $z=1$ and $z=2$.
When we consider the single tracer, $\sigma (f_{\rm NL})$ decreases rapidly as $n_{\rm g}$ increases 
and approach to the CV plateau in the large galaxy number density limit.
Once the galaxy number density is high enough to reach the plateau, the further improvement in
the galaxy number density does not significantly improve constraints on $f_{\rm NL}$.
Even in the case of the two tracer, the plateau appears near $n_{\rm g}\approx 10^{-3}h^3{\rm Mpc}^{-3}$,
presumably because in this region the galaxy number density is high enough to be CV limited but
not enough that the CV cancellation is effective.
Pushing to higher galaxy number density, we found that the reduction of the CV noise
becomes effective and we obtain the stronger constraint on $f_{\rm NL}$.
In our specific survey, the multitracer technique have the potential to constrain on PNG to an accuracy
a factor of $1.4$ better than the single-tracer constraints.
In addition, we found that the further increase in the number of tracers leads to slight improvements
of the constraints.
Hence, in the subsequent analysis, we consider only the case of two mass bins.

Based on the derived analysis tool, we then apply our Fisher matrix analysis to 
future redshift surveys, in which
we will find a large number of galaxies enough that the multitracer technique is effective.
As future representative surveys, we consider the galaxy surveys conducted by 
Euclid and the SKA Phase-2.
We adopt the predicted number density of galaxies as a function of redshift, 
given in Table 3 of Ref.~\cite{Amendola:2016saw} for Euclid and in Table 1 of Ref.~\cite{Bull:2015nra} for the SKA, respectively.
In Fig.~\ref{fig:fNL-b1}, we plot the $1\sigma$ expected marginalized contours in
$(f_{\rm NL}, b_1^{(1)})$ plane.
As for $f_{\rm NL}^{\rm eq}$, the SKA can reach $\sigma (f_{\rm NL}^{\rm eq})=25.1$(1-tracer), 
$23.0$ (2-tracer), which is an improvement by a factor $2$ compared with
the Planck constraint.
The constraint, from Euclid, $\sigma (f_{\rm NL}^{\rm eq})=30.4$ (1-tracer), $30.0$ (2-tracer),
is relatively weaker than the SKA one.
The constraints on $f_{\rm NL}^{\rm orth}$ from Euclid and the SKA are
$\sigma (f_{\rm NL}^{\rm orth})=13.6$ (Euclid, 1-tracer), $13.1$ (Euclid, 2-tracer), 
$12.4$ (SKA, 1-tracer) and $10.2$ (SKA, 2-tracer), respectively. 
The SKA is found to be more advantageous to apply the multitracer technique, 
simply because the low-$z$ source density provided by the SKA is higher than Euclid one.
Therefore, we conclude that the precise measurement of galaxy bispectrum by future galaxy surveys
can probe the non-local type PNG to the level comparable to
or severer than the CMB constraints.

\section{Summary} 

To summarize, we have discussed the potential power of the multitracer technique
for the galaxy bispectrum as a possible probe of the various types of PNG. 
To apply the multitracer technique, we have first derived the formulas for
the monopole mode of RSD for the galaxy bispectrum generalized to multiple tracers.
Performing the Fisher matrix analysis based on the derived formulae, 
we showed that the precise measurement of the galaxy bispectrum with the multitracer technique 
provides the powerful probe of not only the local but also non-local types of PNG without the CV noise.
Particularly, in the region that the galaxy number density is high enough,
even for the case of two tracers, the reduction of the CV noise 
due to the effect of the multiple tracers becomes effective and we obtain 
the stronger constraints on parameters than the single tracer constraints.
Based on these facts, we also found that planned galaxy surveys in the next decade 
indeed have the potential to be competitive with current and future CMB measurements.

In this paper, we have made several simplified assumptions.
We have considered only the tree-level contributions to the galaxy bispectrum
from the gravitational evolution and the primordial bispectrum
with the Kaiser formula.
The higher-order contributions may affect the details of our result,
though generic features are expected to remain the same.
On the other hand, future galaxy surveys will be limited by 
the systematic uncertainties and the CV noises rather than
statistical errors because future surveys will be able to 
probe the huge number of samples.
Hence, we should also address the systematics of future surveys
in more realistic situations.
To take advantage of the multitracer technique, we need to estimate 
the halo mass of each galaxy, which has to be inferred from available observables.
The uncertainty in estimates of the halo mass for individual galaxies
may become important as systematics.
When we consider a future survey that covers a wide area of sky and a significant redshift depth, 
a number of nuisance parameters should be included to model systematic errors.
We hope to come back to these issues in the near future.

\acknowledgments
This work was supported in part by Grant-in-Aid from
the Ministry of Education, Culture, Sports, Science and Technology (MEXT) of Japan,  Nos.~24340048, 26610048, 15H05896, 16H05999 (K.~T.), 15K17659, 15H05888, and 16H01103 (S.~Y.).



\end{document}